# Zwitterions Layer at but Do Not Screen Electrified Interfaces

Muhammad Ghifari Ridwan, Buddha Ratna Shrestha, Nischal Maharjan, Himanshu Mishra[*]

*King Abdullah University of Science and Technology (KAUST), Water Desalination and Reuse Center (WDRC), Biological and Environmental Science & Engineering Division (BESE), Thuwal, 23955-6900, Saudi Arabia*

[*]Corresponding Author: himanshu.mishra@kaust.edu.sa

## ABSTRACT

The role of ionic electrostatics in colloidal processes is well-understood in natural and applied contexts; however, the electrostatic contribution of zwitterions, known to be present in copious amounts in extremophiles, has not been extensively explored. In response, we studied the effects of glycine as a surrogate zwitterion, ion, and osmolyte on the electrostatic forces between negatively charged mica–mica and silica–silica interfaces. Our results reveal that while zwitterions layer at electrified interfaces and contribute to solutions' osmolality, they do not affect at all the surface potentials, the electrostatic surface forces (magnitude and range), and solutions' conductivity across 0.3–30 mM glycine concentration. We infer that the zwitterionic structure imposes an inseparability among the ± charges and that this inseparability prevents the buildup of a counter-charge at interfaces. These elemental experimental results pinpoint how zwitterions enable extremophiles to cope with the osmotic stress without affecting finely tuned electrostatic force balance.

# INTRODUCTION

Ions and zwitterions orchestrate the inner workings of prokaryotic, plant, and animal cells via electromagnetic interactions, thereby giving rise to finely tuned structure–function relationships in proteins, chemical reactions, catalysis, molecular recognition and signaling, and bioenergetics[1-5]. Ion electrostatics also underlies numerous practical matters, including soil water-holding capacity[6], the stability of pharmaceutical and cosmetic formulations[7], water desalination[8] and treatment[9] processes, and nanotriboelectricity harvesting[10]. Interestingly, simple hard ions, such as $Na^+$, $K^+$, $Mg^{2+}$, $Ca^{2+}$, and $Cl^–$, which are ubiquitous in biological systems, retain their electrical charge irrespective of the solution pH or temperature[11]. Therefore, their presence in excess can tilt the finely tuned balance of molecular forces, notably electrostatics[2, 12-14]. The Debye–Hückel model accurately captures the behavior of ions in dilute solution (≤100 mM) on the basis of Maxwell's first law of electromagnetism, $\nabla^2 \phi = -\rho/\varepsilon_0\varepsilon_r$, and the assumption that the Boltzmann statistics accurately describe the clustering of the oppositely charged ions at electrified interfaces, $\rho = \rho_0 e^{-e\psi/k_B T}$ [15]. This Poisson–Boltzmann equation explains why the clustering of ions dramatically decreases the range and magnitude of electrostatic forces between charges and/or charged surfaces, such as within or among proteins and emulsified oil droplets in water[16]. Such a molecular-scale disruption manifests as cytotoxicity[17]; indeed, salt stress precludes the use of seawater for growing food and forces us to exploit limited and dwindling freshwater resources[18]. This severity echoes in "The Rime of the Ancient Mariner": "Water, water everywhere, Nor any drop to drink."[19]. Yet, some lifeforms thrive even in harsh environments, including salty, arid, pressurized, hot, or cold environments, where the amounts of solutes (ions) and solvent (water) can vary dramatically, thereby catastrophically affecting the electrostatics and inducing osmotic imbalance[1, 20, 21]. The current understanding of how life prevails under such extreme conditions via intermolecular and surface forces is far from satisfactory[3, 22-24].

Osmolytes—electrically neutral molecules such as zwitterionic amino acids (e.g., glycine, proline, and alanine), sugars and polyols (e.g., glucose and glycerol), methylamines (e.g., sarcosine, betaine, and trimethylamineoxide), and urea—orchestrate the balancing act[1, 4, 21, 25]. They are observed in high concentrations in a wide variety of extremophiles, including cyanobacteria, fungi, lichens, multicellular algae, vascular plants, insects, and marine invertebrates and pelagic fishes. Researchers have documented the effects of osmolytes and their compensating effects, such as those of betaine and urea, on enzymatic activities.

However, unlike the contributions of hard ions, the contributions of zwitterions to an aquatic solution's ionic strength and surface forces remain unclear. For instance, the zwitterionic contribution to ionic strength has been suggested to be (i) zero[26], (ii) similar to that of 1:1 salts[27], (iii) and similar to that of partially charged molecules[28]. Obviously, these proposed contributions would result in drastically different electrostatic forces[29-31]. Therefore, direct measurements of surface forces as a function of surfaces and solutions are necessary to clarify this matter. Recently, Sivan and coworkers used atomic force microscopy (AFM) to probe the effects of adding betaine (1–3 M) on the forces between silica surfaces immersed in solutions comprising NaCl, KCl, CsCl, and $MgCl_2$ at concentrations less than or equal to 50 mM[32]. They found that, although the addition of hard ions decreased the range and magnitude of electrostatic forces, betaine (1–3 M) increased both the magnitude and range of electrostatics between silica surfaces. These new results suggest that the effect of osmolyte action on surface forces is profound.

Herein, we investigate the effects of glycine, as a surrogate osmolyte, on electrostatic forces between electrified surfaces at biologically relevant concentrations (<0.6 M) to address the following questions:

1. How do zwitterions influence surface forces between electrically charged surfaces in dilute electrolytes?
    (i) Do zwitterions layer at electrified interfaces and screen them, similar to the effect of hard ions?
    (ii) If zwitterions increase/decrease surface forces, is this effect due to their contribution to the solution's ionic strength or their contribution to its dielectric response?
    (iii) If they exert no effect, is the lack of effect attributable to the fortuitous cancellation of their various aforementioned influences?
2. Do zwitterions transition to simple ions if the pH is adjusted to change their charge to, for example, $\pm e$, where *e* is the electronic charge?
    (i) Are the effects of the thus-formed positive and negatively charged ions on surface forces identical?
    (ii) What is the correlation between the surface forces and the electrolytes' osmotic pressure in these systems?

To probe these nested and interrelated questions, we used AFM and a surface force apparatus (SFA) to measure the forces between electrified silica–silica and mica–mica surfaces separated by dilute aqueous solutions (≤30 mM). The results reveal that zwitterions exhibit stronger effects than hard ions as a function of pH and ionic composition; although they enhance solutions' osmolality, they do not influence the electrostatic surface potentials/forces or the screening lengths. In addition, if they are rendered to ionic form by varying the pH, they behave as hard ions.

**RESULTS**

Single-crystal $SiO_2$/Si wafers and freshly cleaved muscovite mica surfaces are ultrasmooth and acquire a negative charge in aqueous solutions depending on the pH/$pK_a$ relationships because of the deprotonation of Si–OH groups[33] and the leaching of $K^+$ ions[15], respectively. These materials therefore serve as rigid substrates for comparing the behaviors of ions and zwitterions at electrified interfaces. Glycine was used as a surrogate osmolyte because it is a common amino acid with the smallest hydrophobic unit. In the pH range 3–9, the majority of glycine molecules in water exist in zwitterionic form; below and above this range, they display net positive and negative charges because of the –$NH_3^+$ and –$COO^-$ groups, respectively[34]. We next used AFM to measure the electrostatic force at the silica–silica interfaces and to probe glycine adsorption onto the electrified mica–water interface. Whereas colloidal probes were used in the former experiments, nanoscale tips were used in the latter experiments (Fig. 1A and Methods). In addition, we used SFA to achieve angstrom-scale resolution between ultrasmooth surfaces while measuring forces and pinpoint electrostatic decay lengths in various solutions[35]. Notably, although the contact area in the colloidal probe experiments was <300 $nm^2$, that in the SFA experiments was ~100 $\mu m^2$ (Fig. 1B and Methods). Results from these complementary techniques are presented in the following subsections.

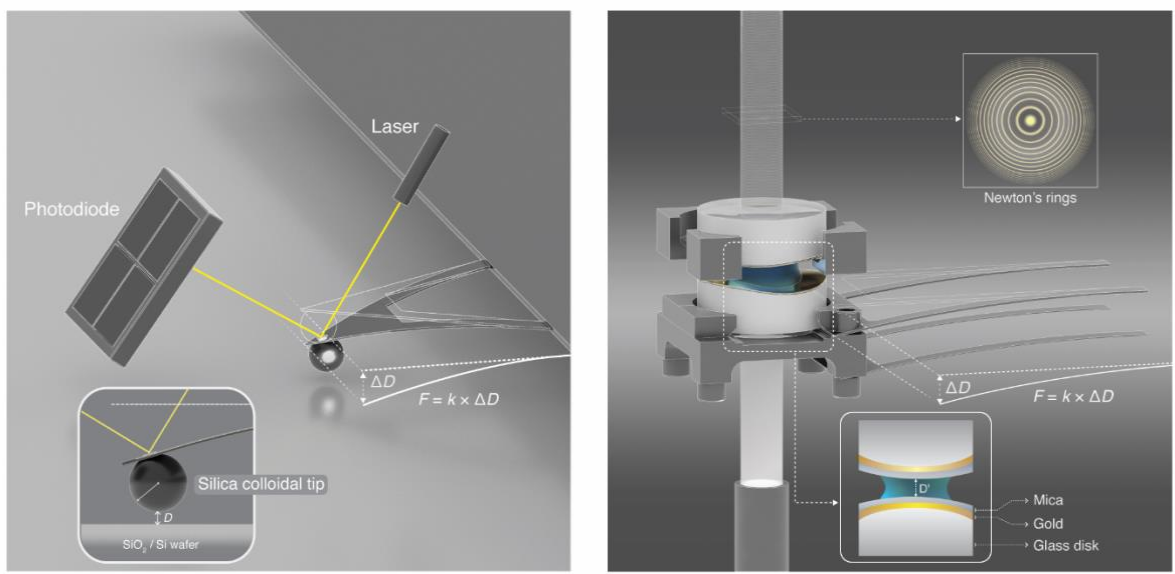

**Figure 1.** (A) Schematics of the atomic force microscopy (AFM) experiments. The relative distance between the AFM cantilever and the surface is determined by the displacement of the laser in the photodiode. The force is then measured on the basis of Hooke's law by detecting the deflection of the cantilever by the prevailing force between the AFM cantilever tip and the surface. The inset shows the contact geometry between a silica colloidal tip and a silica surface. (B) Schematic of the surface force apparatus (SFA). The absolute distance between two surfaces is calculated by interpreting the fringes of equal chromatic order (FECO). The force is then measured on the basis of Hooke's law by considering the difference between the calculated distance and the normal distance (without spring deflection). (Image credits: Heno Hwang, Scientific Illustrator, KAUST)

**Effects of glycine activity and water pH on electrostatic forces between surfaces.**

We conducted AFM experiments (for silica surfaces) and SFA experiments (for mica surfaces) using aqueous solutions of glycine in the concentration range 0.3–30 mM (pH ≈ 6.5) and pure water (pH ≈ 5.7) for comparison. Even though the zwitterionic concentration increased more than 100-fold, no differences were detected in the magnitude of the electrostatic forces and the Debye length in the silica–silica system (Fig. 2A) and the mica–mica system in water (Fig. 2B). That is, the addition of zwitterions did not affect the electrostatics in these systems; the electrostatics were similar to those in pure water. However, when we added hard ions (i.e., KCl (0.01–10 mM) to 3 mM glycine solutions, the force magnitude decreased and the Debye lengths decreased from 45 nm at 0.01 mM to 29 nm at 0.1 mM, 10 nm at 1 mM, and 3 nm at 10 mM. These changes can be explained on the basis of the linearized Poisson–Boltzmann model (see details in the Discussion).

Next, we investigated the effects of the pH of 3 mM glycine solutions in the range 2–12 on surfaces forces for the silica–silica system (Fig. 2C) and the mica–mica system (Fig. 2D). The experimental results revealed that, when $5 < pH < 7$, the Debye length did not change (45 ± 7 nm); however, when pH < 5 or pH > 7, the Debye length decreased systematically as follows: from 29 nm at pH 4.4 to 3 nm at pH 2.1 and from 15 nm at pH 9.8 to 4 nm at pH 11.8. In the Discussion section, we explain these observations based on the linearized Poisson-Boltzmann model and the speciation of glycine (into zwitterions and ions).

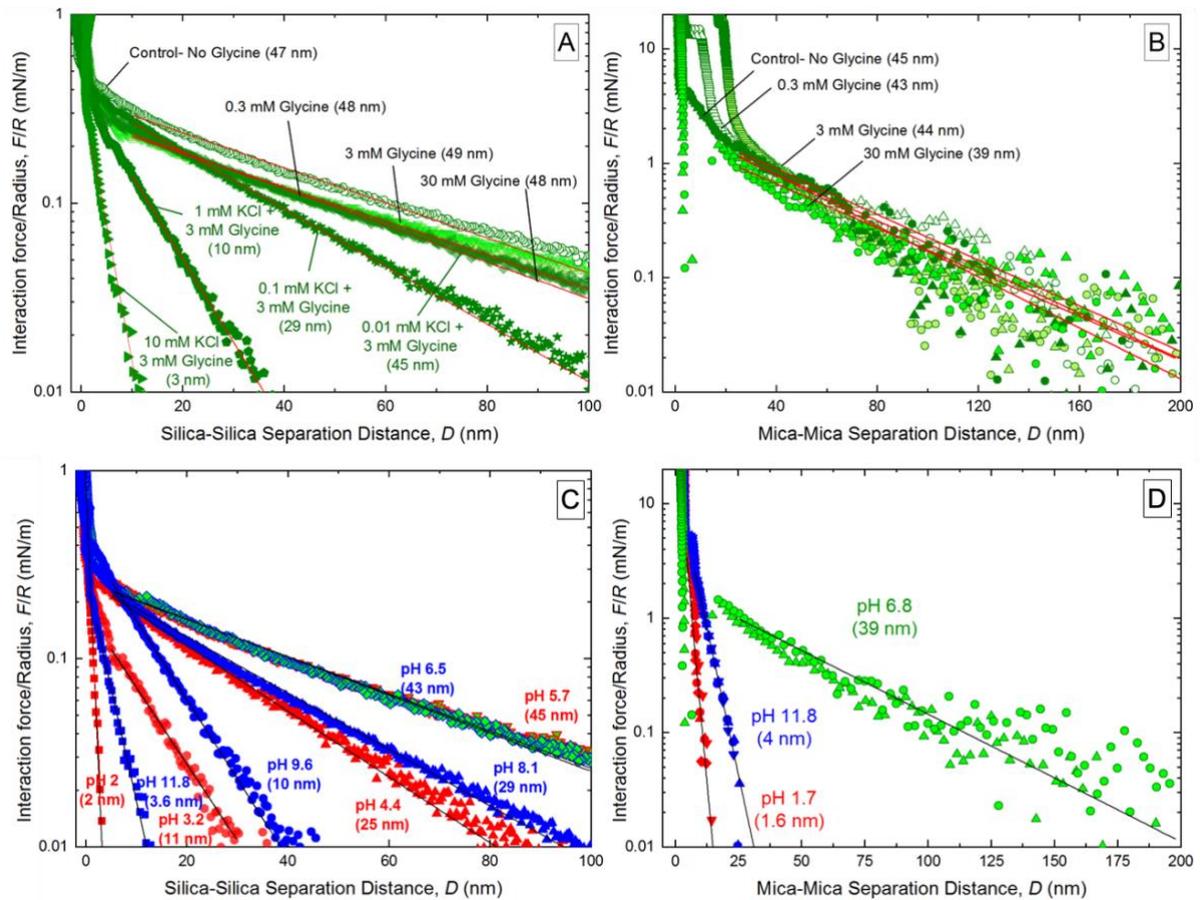

**Figure 2. (A-B)** Effects of the zwitterionic glycine concentration (0.3–30 mM; pH≈ 6.5) on the electrostatic forces between charged surfaces in water with/without KCl. Semi-logarithmic (A) AFM force data normalized by tip radius and (B) SFA force data normalized by the effective radius of curvature of the discs. **(C-D)** Normalized force as a function of the surface separation at different pH values in 3 mM glycine solutions, as measured (C) between silica surfaces by AFM and (D) between mica surfaces by SFA. Under acidic conditions, the zwitterions and positively charged ions formed because protonation of the amine group dominated the chemical speciation of glycine. Under basic conditions, the zwitterions and negatively charged ions formed because deprotonation of the carboxylic group dominated the chemical speciation of glycine. The continuous lines are linear fits whose slope yields the Debye length, $\lambda$, which is listed within parentheses.

**Adsorption of glycine at an electrified interface as a function of water pH.**

We first incubated freshly cleaved mica surfaces in 30 mM glycine solutions at pH values of 1.7, 6.8, and 11.8 for 30 min to achieve chemical equilibrium. We then rinsed the surfaces with deionized water, dried them with flowing $N_2$ gas, and imaged them by AFM (Fig. 3A–C). We found that glycine self-assembled on mica only at pH 6.8, forming a patchy layer (Fig. 3B); no adsorption occurred at pH values of 1.7 and 11.8 (Fig. 3A and C). These results underscore the effects of the form of glycine (i.e., zwitterionic at pH 6.8 and ionic at other pH values) on its adsorption behavior (see details in the Discussion).

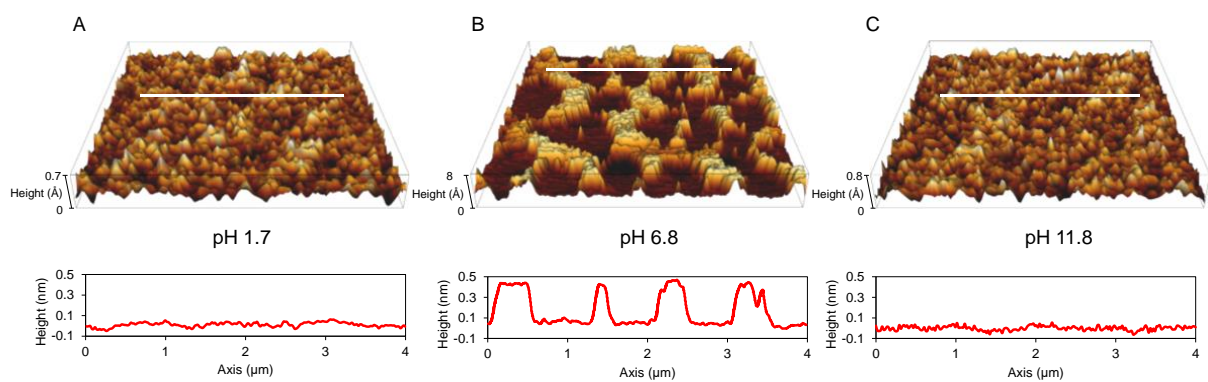

**Figure 3.** 3D AFM images of incubated mica in 30 mM glycine solution at different pH levels. Each image size is $5 \times 5$ µm$^2$. (A) Mica surface after incubation in 30 mM glycine solution at pH 1.7. The asperities' height is less than 0.7 Å. (C) Mica surface after incubation in 30 mM glycine solution at pH 6.8. The asperities' height is 4 Å. (C) Mica surface after incubation in 30 mM glycine solution at pH 11.8. The asperities' height is less than 0.8 Å.

**Electrical conductivity, osmotic pressure, and dielectric constant of glycine solutions.**

We measured the ionic conductivities of glycine solutions as a function of their concentration and solution pH. At pH $7 \pm 0.25$, the ionic conductivities for 0.3 mM, 3 mM, and 30 mM glycine solutions were in the range 30–50 µS/cm (Fig. 4A) and were independent of the solute concentration. In stark contrast, ionic conductivities of 3 mM glycine solutions increased to 250 µS/cm when the solution pH was adjusted to 3 or 9 (Fig. 4B). Specifics of the differences between the anionic and cationic forms are discussed later.

We next quantified the osmolality of aqueous solutions as a function of the glycine concentration, added KCl content, and the pH. The experimental osmolality of the 0.3 mM, 3 mM, and 30 mM glycine solutions was $8.3 \pm 1.7$, $10.0 \pm 0.1$, and $33.5 \pm 4.8$ mOsmol/kg, respectively (Fig. 4C). Similar trends were observed for the glycine solutions with different KCl concentrations; that is, osmolality increased with increasing KCl concentration. In addition, the measured osmolality of the 3 mM glycine solution with a pH in the range 4.4–9.6 was <12 mOsmol/kg, whereas at pH < 4.4 and pH > 9.6, the osmolality of the solution was >12 mOsmol/kg.

Lastly, we measured the dielectric constants of our glycine solutions as a function of their concentrations, added KCl content, and pH at 294 K (Fig. 4D). The results show that the dielectric constants of (i) 0.3–30 mM glycine solutions, (ii) 3 mM glycine solutions containing 0.01–10 mM KCl, and (iii) 3 mM glycine solutions at pH 3.2–11.8 varied within 77.5 ± 0.3 to 79.2 ± 0.5, i.e., within ±1% of the dielectric constant of pure water at 294 K and hence not significant. Curiously, the dielectric constant at pH 1.9 was comparatively lower, with a value of 75.6 ± 0.3. We discuss these results further in the Discussion section.

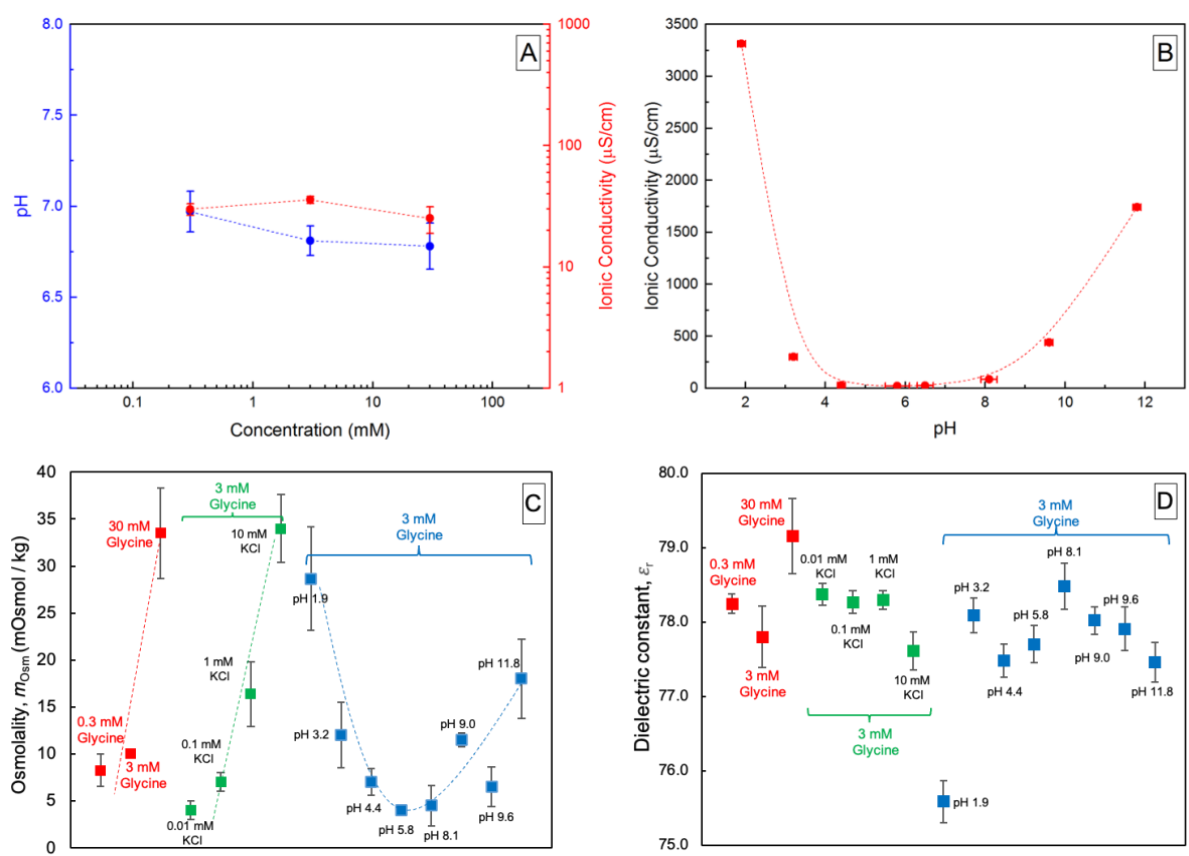

**Figure 4.** (A) Correlations between glycine concentration in water and the solution ionic conductivity (red) and pH (blue). (B) Effects of the solution pH on the ionic conductivities of 3 mM glycine solutions. The osmolality (C) and dielectric constants (D) of glycine solutions as a function of glycine concentration, addition of varying KCl concentrations, and solution pH. (Note: dotted lines/curves in the panels A–C are intended to serve as visual aid only.)

**DISCUSSION**

In this section, we draw upon our results to address the questions posed in the Introduction. **Our key finding is that, in dilute solutions, zwitterions layer/adsorb at electrified interfaces but do not electrostatically screen them.** We explain why the addition of glycine leads to no observable differences in the Debye lengths and surface forces in the concentration range 0.3–30 mM, unlike the case where hard ions are present (Fig. 2A-B).

To facilitate the discussion, we first consider some important formulae related to the range and magnitude of electrostatic surface forces. The first equation is for the Debye length, $\lambda = \left(\frac{2N_A e^2 I}{\varepsilon_o \varepsilon_r k_B T}\right)^{-1/2}$, where $\varepsilon_o \varepsilon_r$ is the permittivity of the medium, $N_A$ is Avogadro's number, $e$ is the electronic charge, $I$ is the ionic strength of the electrolytes, $k_B$ is the Boltzmann constant, and $T$ is the absolute temperature[32]. Ionic strength, in turn, is described by the equation $I = \frac{1}{2}\sum_i C_i z_i^2$, where $C_i$ and $z_i$ denote the concentration and charge of species $i$, respectively. Next, the expression for the normalized surface force investigated in our experiments can be analytically derived as $\frac{F}{R} = \left\{\frac{64\pi\varepsilon_o\varepsilon_r}{\lambda}\left(\frac{k_B T}{e}\right)^2 \tanh^2\left(\frac{e\psi_o}{4kT}\right)\right\}e^{-\frac{D}{\lambda}}$, where $D$ is the separation distance between two surfaces and $\psi_o$ is the surface potential[15].

We now consider the dependence of $\lambda$ on $\varepsilon_r$ and $I$. The negligible variations in the measured surface forces' magnitude and range (Fig. 2A-B) and the dielectric constants of glycine solutions, $\varepsilon_r$, in the range 0.3–30 mM (Fig. 4D) imply that the contribution to $I$ must also be minimal. This argument demonstrates that zwitterions do not contribute to $I$, and hence $\lambda$, in dilute solutions. This result challenges previous proposals for assigning partial/elementary charges[28, 36-42] to zwitterions when estimating their contributions to $I$. Interestingly, the variation in the dielectric constants of 3 mM glycine solutions containing 0.01–10 mM KCl and whose pH values were adjusted in the range 3.2–11.8 is also small. We consider that, in the pH range 4–9, the contributions of hard ions K$^+$, H$^+$, Cl$^-$, and OH$^-$ and zwitterions are opposite: hard ions decrease the dielectric constant, whereas glycine increases it as $\varepsilon_r = 78.5 + \delta \times C_{osmolyte}$ [M], where $\delta = 22.6$/mol [43]. At pH 1.9, the dielectric constant changes substantially, whereas this change is not observed at pH 11.8; this difference warrants further investigation.

Next, we use theory to pinpoint the effects of glycine's zwitterionic and ionic forms on the screening of the electrostatic potential of silica and mica surfaces. To this end, we conducted 100 force-runs in each solution and analyzed the results statistically. Each curve was fitted with an exponential decay function, and pre-exponential factors in the linear (semi-logarithmic force–distance) regime were used to calculate surface potentials at the outer Helmholtz layer[15]. Thus, the calculated surface potentials for silica in 0.3 mM, 3 mM, and 30 mM glycine solutions were $42 \pm 7$ mV (Table 1). By contrast, the addition of 0.01 mM, 0.1 mM, 1 mM, and 10 mM KCl to 3 mM glycine resulted in a dramatic decrease in the calculated surface potential from 40 mV to 38 mV, 25 mV, and 10 mV, respectively. These results demonstrate that zwitterions do not lead to electrostatic screening under dilute conditions. In addition, we repeated these experiments using AFM colloidal probes of different sizes and observed consistent trends (SI Fig. S1).

We here explain the effects of water pH on the Debye lengths and surface forces in glycine solutions, as observed in our experiments with silica–silica (Fig. 2A) and mica–mica systems (Fig. 2B). In the pH range 7.8–11.8, glycine is present in the anionic state because of deprotonation of its $-NH_3^+$ and $-COOH$ groups (SI Figs. S2 and S3); in the pH < 5 range, glycine becomes a cation because of protonation of its $-NH_3$ and $-COO^-$ groups[34]. In its anionic and cationic states, glycine contributes to $I$, which influences $\lambda$ and leads to a decrease in the magnitude and range of surface forces. By contrast, when the pH is in the range 4–9, glycine exists as a zwitterion and does not contribute to $I$, as previously demonstrated; thus, the Debye lengths are unaffected. In addition, the electrical conductivities of solutions containing purely ionic forms of glycine exhibit a conductivity as much as sixfold higher than that of solutions containing the zwitterionic form of glycine. Notably, water's intrinsic ionic strength cannot be neglected in solutions containing purely zwitterionic glycine. Table 1 summarizes the effects of the glycine concentration, added KCl content, and water pH on the electrical conductivity, $I$, $\varepsilon_r$, $m_{osm}$, $\lambda$, repulsive electrostatic forces (at 20 nm), and $\Psi_{silica}$ at various pH values.

**Table 1.** Summary of experimental data and calculated values presented in the present work.

| pH | Composition | Ionic Conductivity (μS/cm) | $I$ (M)[a] | $\varepsilon_r$ | $m_{osm}$ (mOsmol/kg) | $\lambda$ (nm)[b] | $F/R$ (mN/m)[c] | $\Psi_{silica}$ (mV)[b] |
|---|---|---|---|---|---|---|---|---|
| 6.9 ± 0.3 | 0.3 mM glycine | 29.8 ± 3.2 | $6.1 \times 10^{-6}$ | 78.2 ± 0.1 | 33.5 ± 4.8 | 46.1 ± 1.3 | 0.187 | −40.7 ± 3.8 |
| 6.8 ± 0.3 | 3 mM glycine | 35.7 ± 2.4 | $4.8 \times 10^{-6}$ | 77.8 ± 0.4 | 10.0 ± 0.1 | 45.7 ± 7.1 | 0.188 | −41.5 ± 5.0 |
| 6.7 ± 0.2 | 30 mM glycine | 25.1 ± 6.2 | $3.5 \times 10^{-6}$ | 79.2 ± 0.5 | 8.3 ± 1.7 | 48.1 ± 3.6 | 0.186 | −43.4 ± 5.8 |
| 6.4 ± 0.3 | 3 mM glycine + 0.01 mM KCl | 32.4 ± 1.1 | $7.4 \times 10^{-6}$ | 78.4 ± 0.1 | 34.0 ± 3.6 | 43.3 ± 1.6 | 0.188 | −40.4 ± 6.1 |
| 6.2 ± 0.2 | 3 mM glycine + 0.1 mM KCl | 96.3 ± 4.4 | $5.2 \times 10^{-5}$ | 78.3 ± 0.2 | 16.4 ± 3.4 | 27.6 ± 4.1 | 0.185 | −38.6 ± 7.3 |
| 6.2 ± 0.2 | 3 mM glycine + 1 mM KCl | 271.0 ± 0.5 | $5.0 \times 10^{-4}$ | 78.3 ± 0.1 | 7.0 ± 1.0 | 10.1 ± 0.9 | 0.05 | −23.5 ± 8.3 |
| 6.2 ± 0.2 | 3 mM glycine +10 mM KCl | 1461.8 ± 7.9 | $5.0 \times 10^{-3}$ | 77.6 ± 0.3 | 4.0 ± 1.0 | 4.5 ± 0.9 | ~0 | −9.6 ± 5.6 |
| 1.9 ± 0.1 | 3 mM glycine | 3313.1 ± 5.86 | $1.5 \times 10^{-2}$ | 75.6 ± 0.3 | 28.7 ± 5.5 | 1.5 ± 0.5 | ~0 | −4.33 ± 0.6 |
| 3.2 ± 0.1 | 3 mM glycine | 302.4 ± 4.40 | $1.0 \times 10^{-3}$ | 78.1 ± 0.2 | 12.0 ± 3.5 | 10.7 ± 0.9 | 0.03 | −14.9 ± 4.8 |
| 4.4 ± 0.1 | 3 mM glycine | 26.9 ± 5.49 | $6.6 \times 10^{-5}$ | 77.5 ± 0.2 | 7.0 ± 1.4 | 25.6 ± 0.4 | 0.11 | −28.4 ± 3.1 |
| 5.8 ± 0.2 | 3 mM glycine | 21.8 ± 17.16 | $2.6 \times 10^{-6}$ | 77.7 ± 0.2 | 4.0 ± 0.1 | 45.0 ± 2.3 | 0.17 | −41.3 ± 4.9 |
| 6.5 ± 0.3 | 3 mM glycine | 23.4 ± 15.37 | $2.4 \times 10^{-6}$ | 78.5 ± 0.3 | 4.5 ± 2.1 | 43.0 ± 1.6 | 0.17 | −39.8 ± 4.8 |
| 8.1 ± 0.2 | 3 mM glycine | 82.9 ± 7.02 | $9.3 \times 10^{-5}$ | 78.0 ± 0.3 | 11.5 ± 2.1 | 27.3 ± 2.0 | 0.13 | −43.3 ± 4.9 |
| 9.6 ± 0.2 | 3 mM glycine | 439.5 ± 8.96 | $1.5 \times 10^{-3}$ | 77.9 ± 0.3 | 6.5 ± 2.1 | 11.0 ± 0.9 | 0.06 | −23.1 ± 9.3 |
| 11.8 ± 0.1 | 3 mM glycine | 1740.5 ± 14.88 | $9.3 \times 10^{-3}$ | 77.5 ± 0.3 | 18.0 ± 4.2 | 3.6 ± 0.6 | ~0 | −6.9 ± 1.5 |

[a]Zwitterions do not contribute to ionic strength.
[b]Average values from 100 force curves.
[c]Magnitude of surface forces at 20 nm.

We here discuss our mechanistic understanding of the observed similarities and differences in the effects of zwitterions and ions at electrified interfaces. Textbook physics tells us that electrified interfaces repel similarly charged ions and attract counter-ions to form a diffuse electrical double layer (EDL). Consequently, the surface potential (or electrical charge density) perceived outside the EDL is substantially reduced (Fig. 5A). Although zwitterions adsorb onto electrified interfaces (Fig. 3B), they do not affect the net electrical potential (or the surface charge density) because each zwitterionic species comprises an explicit positive and negative charge (Fig. 5B). This inseparability of the + and − charges precludes the accumulation of counterions, thereby obviating screening of electrical field/potential/charge density. Depending on the surface charge, adsorbed zwitterions might orient/distribute in a ± or ∓ alignment with the surface; depending on the surface charge density, a lateral interdigitation, ± ∓ ± ∓ ± ∓, might also be possible along the surface. Nanoscale confinement between charged surfaces of similar/dissimilar surface charge density could further complicate this matter. These aspects should be explored further via complementary experiments and theory. By contrast, when the solution pH is 11.8 (Fig. 3C) and glycine transitions to its anionic form, it is repelled by the negatively charged surfaces. The cationic form (pH 1.7) (Fig. 3A) also fails to adsorb onto the negatively charged surface; it is outcompeted by protons because their small size enables them to better fit into the negatively charged sites[44]. At electrified interfaces, zwitterions form a layer akin to the Helmholtz layer of hard ions; however, their distribution might not follow Boltzmann statistics.

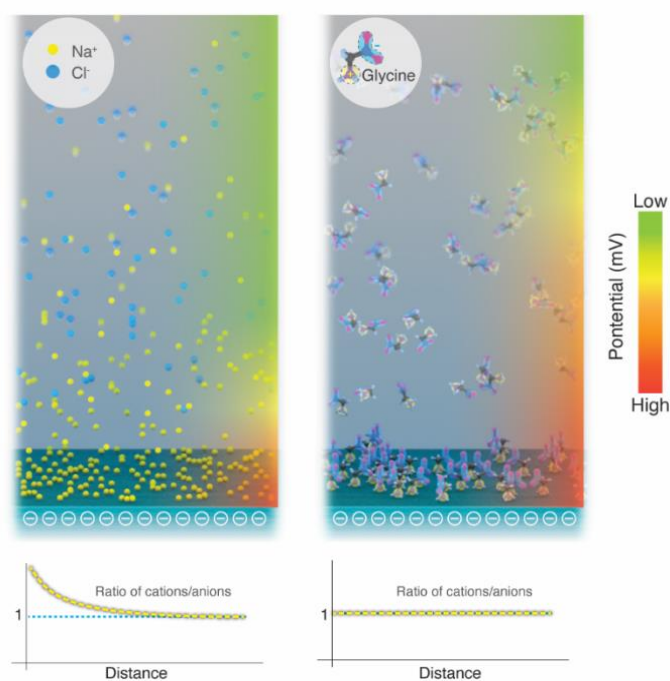

**Figure 5.** Schematic of the ion/molecule distribution on the negatively charged surface in an aqueous solution. (A) Illustration of monovalent ions on the negatively charged surface in an aqueous solution. The inset shows the concentration of monovalent cations and anions as a function of the distance from a negatively charged surface. (B) Illustration of zwitterions on the negatively charged surface in an aqueous solution. The inset illustrates the concentration of cations and anions on a zwitterionic osmolyte as a function of the distance from a negatively charged surface.

Here, we comment on the contribution of glycine speciation on surface forces in concentrated solutions. We remind the reader that, in the pH range 5–7, more than 99.95% of glycine remains in the zwitterion state (SI Fig. S3). Thus, in dilute solutions (e.g., ≤30 mM), the ionic form of glycine is only 0.05% and its contribution to the ionic strength is $1.5 \times 10^{-6}$ M. This contribution is, in fact, lower than the ionic strength of water in equilibrium with atmospheric $CO_2$ (pH 5.6, $5 \times 10^{-6}$ M). Therefore, the effects of speciation are negligible in dilute glycine solutions. By contrast, if the glycine concentration is very high, e.g., 3 M in the pH range 5–7, the speciation into the ionic form would be ~15 mM, which is expected to suppress electrostatics with a Debye length of 2–3 nm. Therefore, our results are consistent with the latest findings on the resurrection of electrostatics at high (1–3 M) osmolyte concentrations. Lastly, we note that, even though zwitterions do not affect ionic strength, surface forces, potentials, or Debye lengths, they do contribute to solutions' osmotic pressure (Fig. 4C). The van't Hoff equation describes this relationship as $\Pi = RT \sum_i n_i C_i$, where $\Pi$ is the osmotic pressure, $R$ is the universal gas constant, $T$ is the absolute temperature, and $n_i$ is the van't Hoff factor. Note that the observed nonlinearity in the osmolality measurements at 0.3 mM is attributed to the instrument's limited accuracy outside its operating range (20–3200 mOsmol/kg).

## CONCLUSION

Our curiosity-driven investigation of zwitterions and ions at electrified surfaces revealed that zwitterions exhibit a rich range of effects depending on the solution pH and concentration. Whereas hard ions such as $K^+$ and $Cl^−$ always contribute to ionic strength, zwitterions do not contribute to ionic strength in dilute solutions. Whereas hard ions adsorb at charged interfaces and screen them, zwitterions form a layer at charged interfaces but do not screen them. Therefore, the magnitudes and ranges of electrostatic surface forces remain unaffected in such solutions. These distinctive behaviors of zwitterions are due to their unusual structure that renders the positive and negative charges inseparable. Thus, zwitterions do not impact electrostatics inside the EDL and the Boltzmann distribution is consequently not relevant to describe their interfacial activity. Zwitterionic surface adsorption would likely depend on the interfacial charge density, the molecular dimensions of zwitterions, and competing effects with other species, e.g., protons; molecular simulations are warranted to probe this further. Our surface force measurements demonstrate how zwitterions, but not hard ions, can maintain the finely tuned balance of electrostatic forces, Debye lengths, and electrical conductivities in dilute solutions with concentrations varying more than 100-fold. Simultaneously, zwitterions contribute to the osmotic pressure in the same manner as hard ions. Therefore, they can facilitate a reliable evolutionary strategy to support life under osmotic stresses. Our report thus provides a surface-forces-based reductionist rationale for the exploitation of zwitterionic osmolytes such as glycine by extremophiles. These findings should also guide the rational design of biocatalysts[45], energy harvesting[46], and nanofluidic devices[47, 48], and beyond[21].

## METHODS

**Materials.** Glycine, potassium chloride (KCl), potassium hydroxide (KOH), and hydrochloric acid (HCl) were purchased from Sigma-Aldrich and were used as-received. Glycine was added to deionized (DI) water from a MilliQ Advantage 10 system (resistivity of 18 MΩ cm, pH 5.7 ± 0.1, and total organic carbon (TOC) ≤ 2 ppm) to prepare solutions with specific concentrations of 0.3 mM, 3 mM, and 30 mM glycine. The solutions were titrated with KOH or HCl to obtain the desired pH. All the experiments were conducted at 21.0 ± 0.5°C. AFM cantilevers with a silica colloidal tip were purchased from Nanoscribe. Si wafers (<100> orientation) with a 2.4 μm-thick thermal oxide layer were purchased from Silicon Valley Microelectronics. Muscovite mica substrates were purchased from S&J Trading.

**Solution pH, ionic conductivity, osmotic pressure, and dielectric constant.** Solutions' pH and conductivity were quantified using a Mettler SevenCompact Duo S213 pH/conductivity benchtop meter. Prior to the measurements, the instrument was calibrated using standard solutions of pH 4, 7, and 10 and solutions with electrical conductivities of 5 μS/cm and 1443 μS/cm. A Vapro 5600 vapor pressure osmometer was used to measure solutions' osmolality after being calibrated with standard solutions with osmolalities of 100 mmol/kg, 290 mmol/kg, and 1000 mmol/kg. The dielectric constants of solutions were measured using an open-ended coaxial probe connected to a frequency vector analyzer (300 kHz to 4.5 GHz). Prior to the measurement, the instrument was calibrated with open, short, load (DI water) calibration.

**Atomic force microscopy (AFM).** A JPK Nanowizard Ultraspeed-II atomic force microscope was used to image glycine adsorbed onto the mica surface and measuring the interaction forces between the silica surfaces in dilute solutions. For the imaging, AFM cantilevers with Sb-doped silica tips (spring constant, $k$ = 2.8 N/m) were used in tapping mode. For the surface force measurement, AFM cantilevers with silica colloidal probes at their tips (tip diameter, $D$ = 15 ± 3 μm; $k$ = 0.32 N/m) were used against a $SiO_2$ (2.4 μm)/Si surface of a <001> Si wafer. The sensitivity and $k$ of the cantilevers were calibrated via the contact-based and thermal noise methods, respectively[49]. The force between the colloidal probe and the substrate was measured by recording the deflection, $\Delta d$, which converted into force using Hooke's law, $F = k\Delta d$. Prior to the measurement, the AFM cantilevers and substrates were cleaned using $O_2$ plasma generated in a Diener Zepto plasma system (process conditions: radio-frequency power = 100 W, pressure = 300 mTorr, $O_2$ flow rate = 16.5 sccm, duration = 3 min).

**Surface force apparatus (SFA).** An SFA-2000 apparatus (SurForce, Santa Barbara, USA) was used to simultaneously measure distances and forces between molecularly smooth mica films in aqueous solutions. To prepare the mica film, one side of them was coated with a 50 nm-thick Au layer; the Au-coated mica films were then glued onto transparent silica discs with a cylindrical face with a radius of curvature, $R$, of 1–2 cm. The Au-coated side was glued, and the pristine mica surface remained exposed. Pairs of mica/Au/glue/disk samples, where the mica was obtained from the same "mother" film to ensure equal thickness, were placed in a cross-cylinder geometry. The distances and forces between the surfaces were determined via white-light multiple-beam interferometry that yielded fringes of equal chromatic order (FECO)[35]. Surfaces were first brought into contact in a dry $N_2$ environment to assess the films' thickness; the surfaces were then separated and approximately 50 µL of an aqueous solution was placed between the samples. The top surface remained fixed, and the bottom surface affixed to a cantilever was driven upward at ~10 nm/s using a motor. Repulsion between the surfaces reduced the approach speed, which bent the cantilever ($k \approx 2$ kN/m); this bending force was recorded to characterize the surface forces.

## DATA AVAILABILITY

The authors declare that all the data supporting the findings of this study are available within the paper and its Supplementary Information. All data available in this work are available from the authors per request.

**Acknowledgments**

HM thanks Mr. Changzi Wang, a student from his course on Aquatic Chemistry (EnSE 202) at KAUST, for bringing this problem to his attention. The authors are indebted to Mrs. Masha Belyi (Research Scientist, Amazon) for creating a Python script for analyzing hundreds of AFM force-distance curves generated in this work to pinpoint trends in Debye lengths and surface potentials. MGR thanks Mr. Mohammad Abbas (KAUST) for discussions on cell physiology; BRS thanks Dr. Bruno Torres (KAUST) for providing colloidal probes for AFM experiments. The co-authors thank Ms. Ana Rouseva (KAUST) and Mr. Paulus Buijs (KAUST) for assisting with osmotic pressure measurements presented in Fig. 5C; Dr. Farizal Hakiki (KAUST) and Prof. Carlos Santamarina (KAUST) for characterizing dielectric responses of solutions presented in Fig. 5D; and Mr. Heno Hwang, KAUST Illustrator, for preparing Figures 1 and 5.

**AUTHOR CONTRIBUTIONS**

HM conceived the project and supervised the research. MGR performed AFM experiments and reproduced BRS's SFA data. BRS performed SFA experiments and reproduced MGR's AFM data. NM performs conductivity and pH measurements. MGR and NM perform osmotic pressure measurements. MGR analyzes the dielectric constant measurement data and osmotic pressure data. MGR wrote the first draft, revised by BRS and NM. HM rewrote and finalized the manuscript.

**COMPETING INTERESTS**

The authors declare no competing interests.

**ADDITIONAL INFORMATION**

**Supplementary information is available for this paper in the submission package.**

# Zwitterions Layer at but Do Not Screen Electrified Interfaces

Muhammad Ghifari Ridwan, Buddha Ratna Shrestha, Nischal Maharjan, Himanshu Mishra[*]

*King Abdullah University of Science and Technology (KAUST), Water Desalination and Reuse Center (WDRC), Biological and Environmental Science & Engineering Division (BESE), Thuwal, 23955-6900, Saudi Arabia*

[*]Corresponding Author: himanshu.mishra@kaust.edu.sa

*Force-distance curves for the silica–silica system obtained via colloidal probe AFM with probe-tips of radius 6.6 μm.*

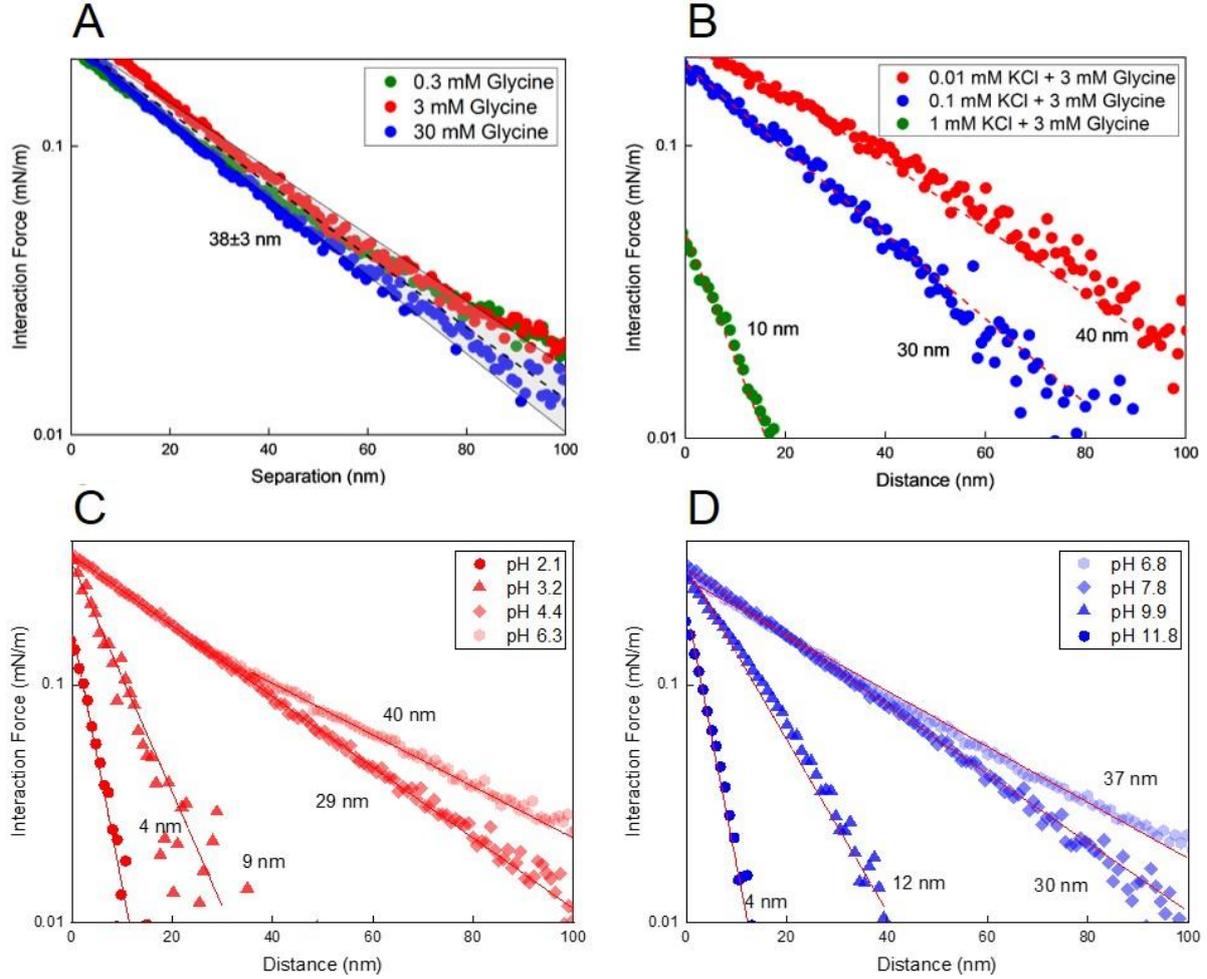

**Figure S1.** Effects of glycine on the screening lengths and the normalized force–distance curves for the silica–silica system probed via AFM using colloidal probes of radius 6.6 µm. (A) Effects of glycine concentration in the range 0.3–30 mM on the measured interaction forces. Dashed black lines are the best fits obtained using the equation $\frac{F}{R} = \left\{\frac{64\pi\varepsilon_0\varepsilon_r}{\lambda}\left(\frac{k_BT}{e}\right)^2 \tanh^2\left(\frac{e\psi_0}{4kT}\right)\right\} e^{-\frac{D}{\lambda}}$, where $\varepsilon_0\varepsilon_r$ is the permittivity of the medium, $e$ is the electronic charge, $k_B$ is the Boltzmann constant, and $T$ is the absolute temperature, $I$ is the ionic strength, $\lambda$ is the Debye length, $D$ is the separation distance between two surfaces and $\psi_0$ is the surface potential[1]. The shaded area represents the errors associated with screening length estimation ($\lambda$). (B) Effects of adding KCl to glycine solutions on the normalized force–distance curves for the silica–silica system probed via colloidal probe AFM. Normalized force as a function of distance between interacting surfaces 3 mM glycine solutions under (C) acidic and (D) basic conditions.

*Speciation of glycine with pH.*

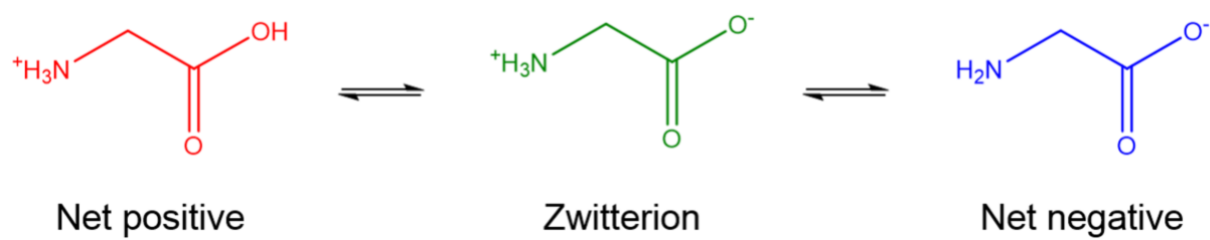

**Figure S2.** The chemical speciation of glycine as a function of solution pH: under acidic conditions, majority of glycine exhibits a net positive charge; under basic conditions, majority of glycine exhibits a net negative charge. At intermediate (near-neutral) pH conditions, glycine is charged but overall neutral. (Please, see Fig. S3 for further details)

*Speciation curves of glycine at various pH.*

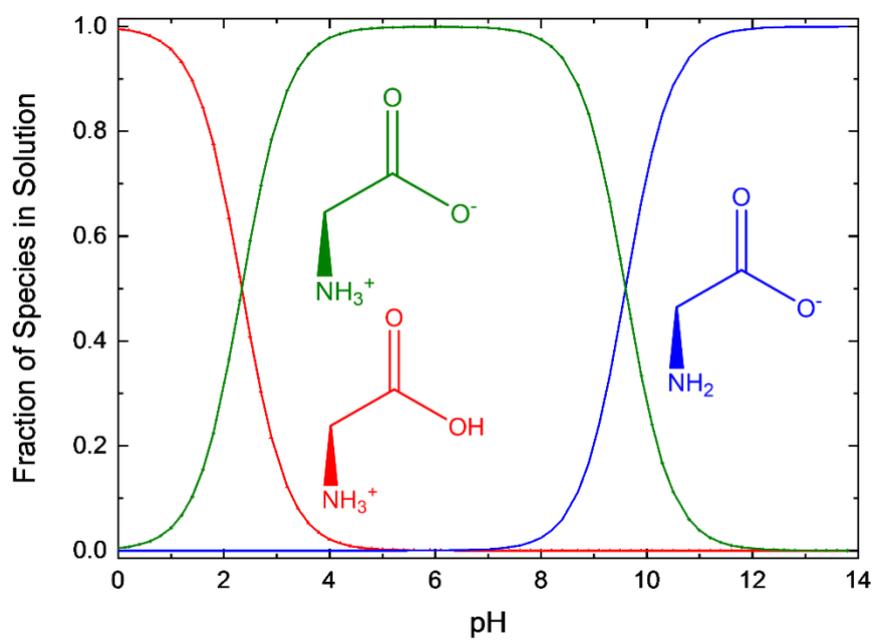

**Figure S3.** The fractions of different states of the amino acid glycine at different pH values.